\documentclass[twocolumn]{revtex4}
\usepackage{graphicx}

\def\a{\alpha}
\def\b{\beta}
\def\ga{\gamma}

\def\s{\sigma}

\def\^#1{\widehat{#1}}

\def\beql#1{\begin{equation} \label{#1}}
\def\beq{\begin{equation}}
\def\eeq{\end{equation}}

\def\<{\langle}
\def\>{\rangle}

\def\({\left(}
\def\){\right)}

\def\[{\left[}
\def\]{\right]}

\def\eqref#1{(\ref{#1})}

\begin{document}

\title{How long does a lockdown need to be?}

\author{Mariano Cadoni\footnote{mariano.cadoni@ca.infn.it}}
\affiliation{Dipartimento di Fisica, Universit\`a di Cagliari,
Cittadella Universitaria, 09042 Monserrato (Italy); \\ {\rm and} \\
INFN, Sezione di Cagliari, 09042 Monserrato (Italy) }

\author{Giuseppe Gaeta\footnote{giuseppe.gaeta@unimi.it}}
\affiliation{
Dipartimento di Matematica, Universit\`a degli Studi
di Milano, via Saldini 50, 20133 Milano (Italy); \\ {\rm and} \\ SMRI,  00058 Santa Marinella (Italy) }



\begin{abstract}
\noindent
Social distancing, often in the form of lockdown, has been adopted by many countries as a way to contrast the spreading of COVID-19. We discuss the temporal aspects of social distancing in contrasting an epidemic diffusion. We argue that a strategy based uniquely on social distancing requires to maintain the relative measures for a very long time, while a more articulate strategy, which also  uses  early detection and prompt isolation, can be both more efficient on reducing the epidemic peak and allow to relax the social distancing measures after a much shorter time. We consider in more detail the situation in Italy, simulating the effect of different strategies through a recently introduced SIR-type epidemiological model. The short answer to the question in the title is: ``it depends on what else you do''.
\end{abstract}

\maketitle

\section{Introduction}

Faced with the first cases of COVID-19, many countries discovered it was spreading much faster than expected, and resorted to some form of more and less extensive lockdown to slow down its spreading.

This was in many cases successful -- depending on how prompt was the decision to implement restrictions -- and indeed it is well known from general properties of epidemic models that \emph{social distancing} has the double effect of lowering the epidemic peak and of slowing down the whole epidemic dynamics. This slowing down allowed to gain precious time to prepare hospitals to face the surge in COVID cases, in particular of those who required hospitalization in Intensive Care Units (ICU).

On the other hand, lockdown has deleterious effects in other regards, in particular social and economical (and also -- in view of how it was implemented in several countries -- sanitary, for what concerns treatment of different pathologies), and poses a severe strain on social structure.

Thus, once the peak of the epidemic wave has passed, it is natural to wonder for how long the restrictive measures should be kept.

In recent papers \cite{Cadoni,Gasir,Gavsb} we have considered the temporal aspects of epidemic dynamic and the consequences in this respect of actions aiming at modifying the main parameters describing the dynamics in SIR-type models. We have shown that actions having the same effect in terms of the height of the peak (maximal number of infectives over the epidemic time-span) and in terms of the total number of people affected by the infection, can have quite different consequences for what concern the temporal development of the dynamics.

In this paper, we will first briefly recall our general results in this direction (these are also discussed in greater detail in a companion paper \cite{CGD}), then pass to consider more specifically a SIR-type model which was quite successful in describing the epidemic data for Italy, and within this model discuss what would be the expected outcomes, in particular for what concerns the time for which the lockdown should be maintained, depending on different strategies which could be implemented now, when all seems to indicate that the peak is -- at least with the presently applied restrictive measures -- behind us.

It should be stressed that these results, albeit applying to a model which so far fits quite well the epidemiological data, are merely indicative and not predictive. Any model describing a population with average characteristics, thus in terms of ``equivalent'' individuals, is clearly too rough to make any sensible prediction. This is even more true in the case of the ongoing COVID epidemics, which affects in quite different ways different geographical regions on the one hand, and people with different sex, age, other pathologies, and so on.

Despite this, the \emph{qualitative} indications of the model are quite clear and, in our opinion, also quite reliable. These are that contrasting the COVID epidemic \emph{only} by social distancing means slowing down the dynamics to a degree which is incompatible with other needs, while a strategy combining social distancing with other actions, in particular \emph{contact tracing, early detection and prompt isolation} of new infectives can reduce even more significantly the epidemic peak and the total number of infected while keeping the period in which social distancing and lockdown are necessary to a reasonable length.

This is no surprise, and actually just confirms in terms of a mathematical model what has been preached theoretically, and implemented on the field in a most effective way, by the Padua group working on the first burst of infections in Veneto \cite{Crisanti}, and later on inspiring the regional strategy for COVID. This strategy was specially successful so far, as shown by the numbers and in particular by the limited mortality in this region.

\section{The SIR model}
\label{sec:SIR}

The classical SIR model \cite{KMK,Murray,Heth,Edel,Britton} concerns averaged equations for a population of ``equivalent'' individuals (thus in Physics'  language it is a mean field theory); each of them can be in three states, i.e. being Susceptible (of infection), Infected and Infective, and Removed (from the infective dynamic).

The populations of these three classes are denoted as $S(t)$, $I(t)$ and $R(t)$ respectively, and the dynamic is defined by the equations
\begin{eqnarray}
dS/dt &=& - \, \a \, S \, I \nonumber \\
dI/dt &=& \a \, S \, I \ - \ \b \, I \label{eq:sir} \\
dR/dt &=& \b \, I \ . \nonumber \end{eqnarray}
where $\a$ and $\b$ are constant parameters, discussed in a moment.

Note that the third equation amounts to a direct integration, $ R(t)  = R(t_0) + \b \int_{t_0}^t I(y ) d y$.
Moreover, the total population $N$ is constant in time (people dying for the considered illness are considered as removed); this makes sense when considering a limited span of time.

The SIR model is described in any textbook on Mathematical Biology, see e.g. \cite{Murray,Edel,Britton}. Here we just recall that the parameter $\a$ corresponds to a \emph{contact rate}, while the parameter $\b$ is a \emph{removal rate}.

Thus social distancing measures work on the reduction of $\a$, while early detection campaigns work on the increase of $\b$.

\subsection{Epidemic threshold, epidemic peak}

Some simple but relevant consequences follow immediately form the SIR equations \eqref{eq:sir}. First of all, it is clear that $I(t)$ grows if and only if
\beql{eq:ga} S(t) \ > \ \frac{\b}{\a} \ := \ \ga \ ; \eeq
for this reason $\ga$ is also known as the \emph{epidemic threshold}. An epidemic can start only if $S(t_0) > \ga$, and it stops spontaneously when $S(t)$ gets below $\ga$.

An equivalent way to describe  the epidemic threshold is to introduce the \emph{reproduction number}
$\rho(t)$ :
\beql{eq:rho(t)} \rho(t) \ = \frac{S(t)}{\gamma}, \eeq
which gives  the expected  new  infections generated  by  a single infection.  The epidemic starts if $\rho(t_0)>1$  (this is the \emph{basic} reproduction number, usually denoted $R_0$),  $I(t)$ attains its peak  value $I_*$ at $t=t_*$  when $\rho(t_*)=1$.  Containment strategies  aim, by reducing $\a$ and/or by rising $\beta$, to reach  $\rho<1$, thus stopping the epidemic.

Also, considering the equations for $S$ and $I$ we easily get a relation between these quantities; moreover we know that $I$ reaches its maximum $I_*$ when $S=\ga$; considering moreover that in the COVID the whole population is initially susceptible to be infected (as far as we know there is no natural immunity), and that the number of infectives at the beginning of the epidemic is negligible compared to the whole population (this approximation will always be used from now on), this turns out to be
  \beql{eq:I*} I_* \ = \ N \ - \ \ga \ - \ \ga \ \log (N/\ga) \ . \eeq
(Needless to say, this expression applies if and only if $N>\ga$ respectively: if $N < \ga$ there is no epidemic.)

The relation between $I$ and $S$ also characterizes the number $R_\infty$ of individuals going through the infected state over the whole epidemic period; the number  $S_\infty$ of those never being in contact with the pathogen is the (lower) root of the equation
\beql{eq:Sinf} (S_0 - S_\infty) \ = \ \ga \ \log (S_0 / S_\infty) \ . \eeq
This transcendental equation cannot be solved in closed form, but it is obvious that the solution will depend only on $\ga$. The number of individuals having gone through infection is of course
\beql{eq:Rinf} R_\infty \ = \ N \ - \ S_\infty \ . \eeq

\subsection{Time to epidemic peak}

The main quantity characterizing the time-span of the epidemic is the \emph{time of occurrence} $t_*$  of the peak. The value of $t_*$ depends on the parameters  $\a$ and $\b$, not just on their ratio, so that containment measures (aimed to reduce $I_*$ and $R_\infty$) do in general have the effect of increasing  $t_*$. More precisely,  $t_*$  can be written in terms  of the parameter $\a$ and $\b$  (see Ref. \cite{Cadoni} for details) as
\begin{eqnarray}
\label{eq:pt}
t_* &=& \int_{0}^{\tau_*} \frac{1}{I_0+S_0-S_0e^{-\a \tau'}-\beta \tau'} \, d \tau' \ , \nonumber \\ \tau_* &=& \frac{1}{\a} \, \log \left( \frac{S_0}{\gamma} \right) \ .
\end{eqnarray}
In the previous expression the integral has to be evaluated numerically. An analytic  expression for $t_*$ can be found using an approximate  solution.     In order to do this we  consider the relation between $S$ and $R$; from the first and third equation in \eqref{eq:sir} we easily get
$$ S \ = \ S_0 \ \exp \[ - \, \(\frac{R \, - \, R_0}{\ga} \) \] \ . $$
Using this and $I(t) = N - S(t) -R(t)$, we can reduce to consider a single ODE, say for $R(t)$. This is written as \beql{eq:R(t)} \frac{dR}{dt} \ = \ \b \ \[ N \ - \ S_0 \,e^{- (R - R_0)/\ga } - R \] \ . \eeq
This is a transcendental equation, and cannot be solved in closed form. It can of course always (and rather easily) be solved numerically.

For $(R-R_0) \ll \ga$, we can expand the exponential in \eqref{eq:R(t)} in a Taylor series, and truncate it at order two. This produces a quadratic equation, which is promptly solved in general terms (see e.g. Sect.~10.2 in \cite{Murray}), yielding
\beql{eq:Rsol} R(t) \, = \, R_0 \, + \, \frac{\a^2}{S_0} \, \[ \( \frac{S_0}{\ga} \, - \, 1 \) \, + \, \kappa \, \tanh \( \frac{\b \,\kappa \, t}{2} \, - \, \phi \) \] \, , \eeq
with constants $\kappa$ and $\phi$ given (for $I_0 \simeq 0$) by
\beq
\kappa \, = \, \( \frac{S_0}{\ga} \, - \, 1 \) \ , \ \
\phi \, = \,  \frac{1}{\kappa} \, \mathrm{arctanh} \[ \frac{S_0}{\ga} \, - \, 1 \] \ . \eeq

As $dR/dt = \b I(t)$, the epidemic peak corresponds to the maximum of $R'(t)$. The solution \eqref{eq:Rsol} allows to compute the time $t_*$ at which this is attained in a straightforward manner; it results
\beq t \ = \ t_* \ = \ \frac{2 \ \phi}{\b \ \kappa} \ .  \eeq
Recalling now the expressions for $\kappa$ and $\phi$, we get that
\beql{eq:t*} t_* \ = \ \frac{2}{\b} \ \frac{\mathrm{arctanh} \( \frac{S_0}{\ga} \, - \, 1 \)}{\( \frac{S_0}{\ga} \, - \, 1 \)^2}  \ . \eeq

This shows that $t_*$ is inversely proportional to $\b$. This relation is not surprising, as $\b$ is the inverse of a time (the removal time). In view of this remark, one has to expect that the inverse proportionality holds also without the assumption $(R - R_0) \ll \ga$. This is indeed the case; showing this goes through use of the scaling properties of the SIR equations and of the integral in  Eq.~\eqref{eq:pt} when one reduces  $\a$ by a factor $\lambda>1$, i.e  $\a\to \a/\lambda$   and  increases $\b$ by the same factor $\b\to \lambda\b$.
While  under these transformations   $I_*$  and $R_\infty$  attain the  same value,  because they do not depend on $\a$ and $\b$ separately but only on their ratio  $\ga$, the value of   $t_*$ is  reduced by a factor $1/\lambda$. (For a detailed  discussion we refer to previous works of ours \cite{Cadoni, Gasir}.)

This means that containment  measures that increase $\b$, e.g. based on tracing and removal of infectives,  have a different effect than -- and in some circumstances an advantage over -- those  that reduce $\a$, e.g based on social distancing or lockdown;  and this in particular for what concerns the timescale of the epidemic.  If  by increasing $\b$ we manage to bring $\rho$ below the threshold we simply stop the epidemic, but even if we do not go so far, we can still  reduce the size of an epidemic keeping under control its timescale.

 This is an advantage when the sanitary system can cope with the epidemic, in that in this way the restrictive measures do not have to be implemented for too long.  The situation is of course different if and when the sanitary system is overwhelmed by the epidemic (as it happened in the first phase of the COVID-19 diffusion in many countries or regions, also as a result of its unexpectedly fast spreading \cite{GR0}). In that situation slowing down the increase of the number of patients is an essential feature, and in this sense social distancing is an essential tool \footnote{Beside, being implemented in a simpler way and not requiring to set up a support organization which can not be improvised.}.

\section{Epidemic management in the SIR framework}
\label{sec:episir}

We have so far supposed that our problem was to analyze the behavior of the SIR system for given control parameters $\a ,\b$ and given initial conditions.

When we are faced to a real epidemic, as the ongoing COVID one, any State will try to manage it, i.e. reduce its effects. When we analyze the situation in terms of the SIR model, this means acting on the parameters $\a$ and/or $\b$.

As discussed above, the strategy are based on two types of actions, i.e. \emph{social distancing}, which in some countries took the extreme form of a lockdown, and \emph{early detection} (followed of course by prompt isolation of infectives); these impact respectively on the $\a$ and on the $\b$ parameter.

From the  previous  discussion it follows   immediately that acting on $\a$ and/or on $\b$ -- in particular reducing $\a$, as in social distancing action -- we can reduce the height of the epidemic peak $I_*$ and simultaneously \emph{increase}  $t_*$, i.e \emph{slow} down  the epidemic dynamic.

The formula \eqref{eq:I*} shows immediately that the height of the epidemic peak depends only on $\ga$, i.e. on the ratio $\b/\a$ of these two parameters. On the other hand, Eq.~\eqref{eq:t*} shows that the time development of the epidemic does \emph{not} depend only on $\ga$, but depends on $\b$.

In particular, we have seen that $t_*$ is proportional (for given $\ga$) to $\b^{-1}$, i.e. decreases as $\b$ increase. But keeping $\ga$ constant means that if $\b$ decreases then $\a$ also decreases, and \emph{viceversa}. Thus a \emph{decrease} in $\a$ at $\ga$ constant corresponds to an \emph{increase} of $t_*$, and more generally to a \emph{slowing down of the epidemic dynamic}.

This is well known, and indeed one of the reasons for the lockdown is to slow down the epidemic increase so to have time to prepare for the epidemic wave, e.g. in terms of Hospital -- or ICU -- capacity, or in terms of Individual Protection Devices (IPD) stocking.

The problem is that social distancing measures -- particularly when they take the form of a lockdown -- are extremely costly in economical and social terms. Thus, once the first epidemic wave has passed and the Hospital system has been reinforced, it is essential to be able to conclude the lockdown in reasonably short times \footnote{Here it should be stressed  once again that albeit in this occasion ``social distancing'' (and hence, in terms of the SIR model, a reduction of $\a$) was considered as equivalent to ``lockdown'', these concepts are \emph{not} the same: e.g., virus transmission rate can be lowered by generalized proper use of IPD, which is a form of social distancing.}.

On the other hand, the main lesson which can be drawn from the results of the previous section is that containment measures which increase $\b$, such as tracing  and removal of infected  individuals, are  in many cases, i.e. except when slowing down the pace of the epidemics is essential due to unpreparedness of the Health System, more suitable for fighting  epidemics because they allow to contain the number of infected people without expanding too  much  the time over which the epidemic is active.

\begin{figure}
    \centering
    \begin{tabular}{cc}
  \includegraphics[width=120pt]{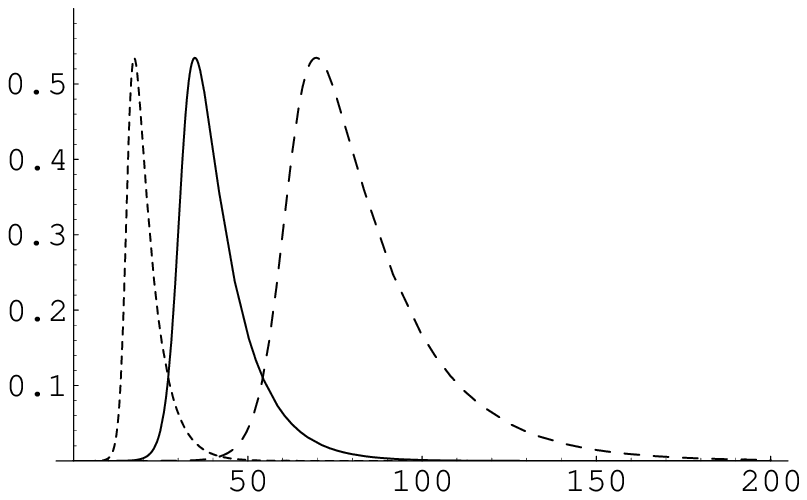}&
  \includegraphics[width=120pt]{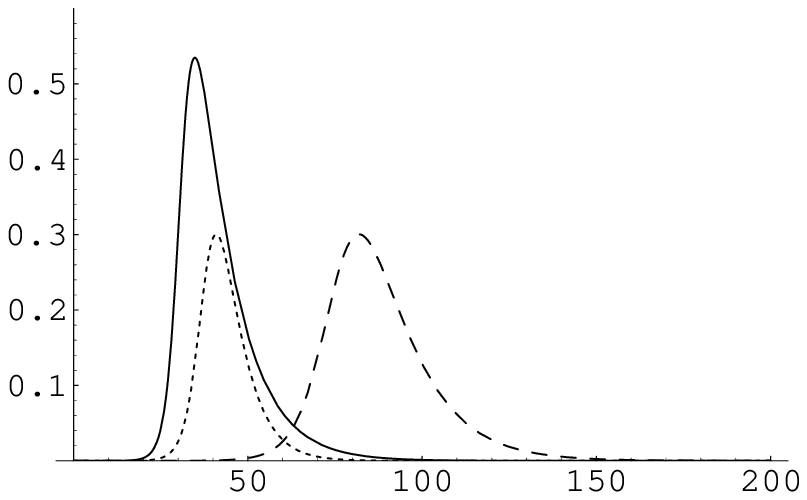}\\
  (A) & (B) \end{tabular}
  \caption{Left side plot (A): Effect of varying parameters in the SIR model while keeping $\ga$ constant. We have considered a population $N=6*10^7$ and integrated the SIR equations with initial conditions $I_0 =10$, $R_0=0$. Setting $\a_0 = 10^{-8}$, $\b_0 =10^{-1} \mathrm{d}^{-1}$, the runs were with $\a = \a_0$, $\b=\b_0$ (solid curve); $\a = \a_0/2$, $\b=\b_0/2$ (dashed curve); and $\a = 2 \a_0$, $\b= 2 \b_0$ (dotted  curve). The curves yield the value of $I(t)/N$, time being measured in days. The epidemic peak reaches the same level, with a rather different dynamics. Right side plot (B): Contrasting the epidemic through different strategies. We plot $I(t)/N$ for the same system and initial conditions as in plot (A). Now we consider $\a = \a_0$, $\b=\b_0$ (solid curve); $\a = \a_0/2$, $\b=\b_0$ (dashed curve); and $\a = \a_0$, $\b= 2 \b_0$ (dotted  curve). Actions reducing $\ga$ by the same factor through action on the different parameters produce the same epidemic peak level, but with a substantially different dynamics.}\label{fig:sir1}
\end{figure}

We will thus discuss for how long social distancing measures can be needed. In this respect, it is convenient to look at Figure \ref{fig:sir1}, where we show the effect of varying $\a$ ad $\b$ in such a way that $\ga$ is constant; and we also show the consequences of raising $\ga$ by the same factor (not sufficient to eradicate the epidemic) through action on the different parameters.

It is apparent from Fig.~\ref{fig:sir1} (right panel) that the price to pay for a strategy based \emph{uniquely} on social distancing measures -- in whatever way they are implemented -- is that these should be maintained for a very long time. On the other hand, a strategy based on early detection and prompt isolation alone reduces the peak without slowing down the dynamics. This feature can of course be a positive or negative one, depending on how ready is the Health System to face the epidemic wave.

\begin{figure}
\centering
  \includegraphics[width=200pt]{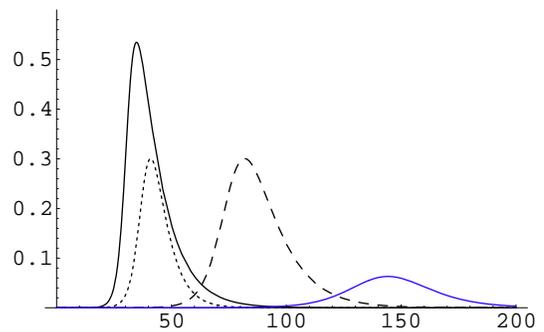}\\
  \caption{Combining action on the $\a$ and the $\b$ parameters. Black curves are as in Fig.~\ref{fig:sir1} (right panel), while the blue curve corresponds to $\a= \a_0/2$, $\b = 2 \b_0$.}\label{fig:sir3}
\end{figure}

In real situations, of course, one should act by combining both strategies; different mix of these will produce different time development, and it is not obvious that the choice should be necessarily for the strategy allowing a greater reduction of the epidemic peak, once all kind of sanitary, economic and social considerations are taken into account. This is shown in Fig.~\ref{fig:sir3}, where it appears that the strategy combining full action on both parameters reduces greatly the level of the epidemic peak, but also makes the epidemic running for a very long time: this may be convenient or not convenient depending on factors which cannot be taken into account by the purely epidemic model, but depends on the  hospital and ICU capacities and on the cost of the lockdown.

For example, we can consider that strict measures should be in effect until the level of infectives descends below a fraction $10^{-4}$ of the population; we denote as $t_s$ the time at which this level is reached. What this means in terms of the duration of the intervention in the different framework considered in Fig.~\ref{fig:sir3} is summarized in Table \ref{tab:sir}.

Note also that here we considered a given initial time of the epidemic and suppose the restrictive measures are maintained until the situation improves enough. In real situations, the restrictive measures will not be set as soon as the epidemic reaches the country, but only after the number of infectives raises above some alert threshold, call this time $t_a$; this has indeed been the pattern in most countries, with some notable exceptions (such as Greece and New Zealand -- they were indeed very lightly struck by COVID). Thus using $t_s$ as an estimate of the length of lockdown leads to overestimating it, and the duration of the measures is better measured by $\tau = t_s - t_a$. This is also considered in Table \ref{tab:sir}, assuming the alert level is the same as the safety one, i.e. $I(t_a) = I(t_s) = 10^{-4} * N$.

In Table \ref{tab:sir} we also give the value of $R_\infty/N$; this shows clearly that in these simulations we do \emph{not} have $R/\ga \ll 1$ (recall $N>\ga$), hence we can not use the approximation leading to \eqref{eq:t*}. In fact, e.g. in Fig.~\ref{fig:sir3} it appears that raising $\b$ does lead to a (small) delay in the epidemic peak, contrary to what is predicted by the ``small epidemic'' formula \eqref{eq:t*}.

\begin{table}
  \centering
  \bigskip
  \begin{tabular}{|c|c|c||l|r||r|r||r||}
  \hline
  $\a/\a_0$ & $\b/\b_0$ & $\ga/\ga_0$ & $I_*/N$ & $t_*$ & $t_s$ & $\tau$ & $R_\infty /N$  \\
  \hline
  1 & 1 & 1 & 0.535 & 35 & 125 & 113 & 0.997 \\
  1/2 & 1/2 & 1 & 0.535 & 70 & 251 & 226 & 0.997 \\
  2 & 2 & 1 & 0.535 & 17 & 63 & 56 & 0.997 \\
  \hline
  1 & 2 & 2 & 0.300 & 41 & 94 & 78 & 0.940 \\
  1/2 & 1 & 2 & 0.300 & 82 & 189 & 157 & 0.940 \\
  1/2 & 2 & 4 & 0.063 & 144 & 246 & 182 & 0.583 \\
  \hline
  \end{tabular}
  \caption{SIR model. Height and timing of the epidemic peak $I_*= I(t_*)$ and time for reaching the ``safe'' level ($I(t_s) = 10^{-4} S_0$), together with duration of the interval $\tau = t_s -t_a$ and the fraction $R_\infty / N$ of individuals having gone through infection, for the different combinations of parameters considered in the numerical runs of Fig.~\ref{fig:sir1} and \ref{fig:sir3}. See text. Note that $R_\infty/N$ depends only on $\ga$, as guaranteed by Eqs.~\eqref{eq:Sinf}, \eqref{eq:Rinf}.}\label{tab:sir}
\end{table}

\section{The A-SIR model}
\label{sec:asir}

The SIR model has a weak point when it comes to modeling the COVID epidemic. That is, it does not take into account the presence of a \emph{large set of asymptomatic infectives}.

In order to take this into account, we have recently proposed a variant of the SIR model, called A-SIR (with the ``A'' standing of course for ``asymptomatic''), see \cite{Gasir}. In this, there are two classes of infected and infective individuals, i.e. symptomatic ones $I$ and asymptomatic ones $J$. Correspondingly, there are two classes of removed, $R$ for those who are isolated -- and eventually dead or recovered -- after displaying symptoms, and $U$ for those who are not detected as infective and hence are removed only when naturally healing.

The key point is that the mechanism of removal is different for the two classes, and hence so is the removal rate. We denote as $\b$ the removal rate for infectives with symptoms; this corresponds to isolation, with a mean time $\b^{-1}$ from infection to isolation. We denote by $\eta$ the removal rate for asymptomatic infectives, with a mean time $\eta^{-1}$ from infection to healing.

If an individual is infected, it has a probability $\xi \in (0,1)$ to develop symptoms. It is assumed that all infective with symptoms are detected and registered by the Health System; as for asymptomatic ones, only a fraction of them is actually detected, and in this case isolated once detected.

Within this description, $\xi$ is a biological constant \footnote{We recall our description applies to ``average'' quantities; it is obvious that in a finer description $\xi$ would be depending on the characteristics of the individual who is infected, e.g. his/her age or general health state.}; on the other hand a campaign for detecting asymptomatic infectives -- e.g. by mass random testing -- would result in a reduction of the average time $\eta^{-1}$, i.e. raising $\eta$. On the other hand, a campaign for \emph{tracing contacts} of registered infectives would result in \emph{early isolation} of would-be new infectives -- both symptomatic and asymptomatic -- and hence in a raising of both $\b$ and $\eta$.

The A-SIR model is described by the equations
\begin{eqnarray}
dS/dt &=& - \, \a \ S \, (I + J ) \nonumber \\
dI/dt &=& \a \, \xi \, S \, (I +J) \ - \ \b \, I \nonumber \\
dJ/dt &=& \a \, (1-\xi) \, S \, (I +J) \ - \ \eta \, J \label{eq:asir} \\
dR/dt &=& \b \, I \nonumber \\
dU/dt &=& \eta \, J \ . \nonumber \end{eqnarray}
Note that the last two equations amount to integrals, i.e. are solved by $R(t) = R_0 + \b \int_0^t I(y) d y$, $U(t) = U_0 + \eta \int_0^t J(y) \ d y$. Moreover, the total population $N = S+I+J+R+U$ is constant.

Unfortunately, for this model no analytical results are available, and one can only resort to numerical simulations. On the other hand, these show that the A-SIR model can account quite well for the COVID epidemic in Italy \cite{Gasir}, i.e. in the only case where it has been tested against real epidemiological data.

In the following it will be convenient to consider also the total number of infectives; this will be denoted as
\beql{eq:K(t)} K(t) \ := \ I(t) \ + \ J(t) \ . \eeq

\section{Epidemic management in the A-SIR framework}
\label{sec:epiasir}

When working in the SIR framework, we have seen that different strategies resulting in the same variation of $\ga$ -- and hence of the epidemic peak and the total number of people going through infection --  yield quite different results in terms of the temporal development of the epidemic. It is quite natural to expect that the same happens in the frame of the A-SIR model.

Now, beside $\a$ and $\b$, we have a third parameter $\eta$. We can thus act on our system in three independent ways, i.e. by \emph{social distancing} (reducing $\a$), by \emph{early isolation} of symptomatic infectives (raising $\b$), and by \emph{detection of asymptomatic}, and of course their isolation (raising $\eta$).

It should be stressed that while implementation of the first strategy is rather clear, at least in principles (things are less clear when other, e.g. economical and social, considerations come into play), for the other two strategies the implementation is less clear, and in practice consists of two intertwined actions: \emph{tracing contacts} of known infectives, and \emph{large scale testing}. To put things in a simple way, \emph{actions on $\b$ and on $\eta$ go usually, in practice, together}. Thus we should essentially distinguish between social distancing on the one side, and other tools on the other side.

\begin{figure}
\centering
  \includegraphics[width=200pt]{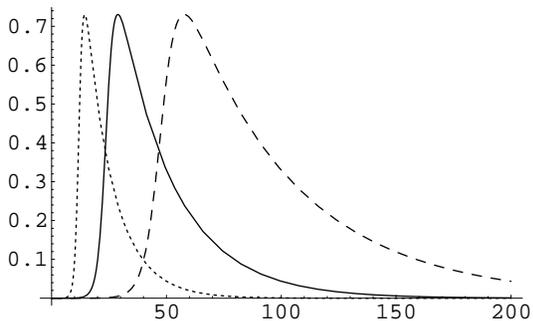}\\
  \caption{Effect of varying parameters in the A-SIR model while keeping $\ga$ constant. We have considered a population $N=6*10^7$ and integrated the A-SIR equations with $\xi = 0.1$ and initial conditions $I_0 =10$, $J_0 = 90$, $R_0=U_0=0$. Setting $\a_0 = 10^{-8}$, $\b_0 =10^{-1} \mathrm{d}^{-1}$, $\eta_0 =4*10^{-2} \mathrm{d}^{-1}$. The runs were with $\a = \a_0$, $\b=\b_0$, $\eta=\eta_0$ (solid curve); $\a = \a_0/2$, $\b=\b_0/2$, $\eta=\eta_0/2$ (dashed curve); and $\a = 2 \a_0$, $\b= 2 \b_0$, $\eta= 2 \eta_0$ (dotted  curve). The curves yield the value of $K(t)/N$, see \eqref{eq:K(t)}, time being measured in days; the plots for $I(t)/N$ would look similar, with a different scale. The epidemic peak reaches the same level, with a rather different dynamics.}\label{fig:asir1}
\end{figure}

\begin{figure}
\centering
  \includegraphics[width=200pt]{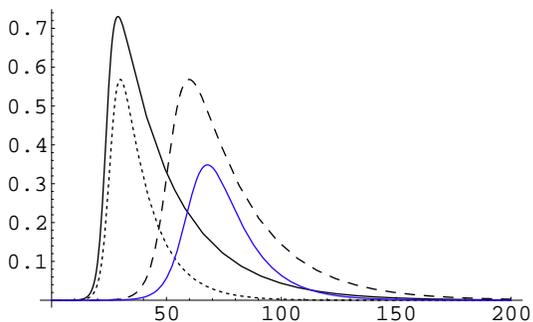}\\
  \caption{Contrasting the epidemic through different strategies. We plot $K(t)/N$ for the same system and initial conditions as in Fig.~\ref{fig:sir1}; the plots for $I(t)/N$ would look similar, with a different scale. Now we consider $\a = \a_0$, $\b=\b_0$, $\eta=\eta_0$ (solid curve, black); $\a = \a_0/2$, $\b=\b_0$, $\eta=\eta_0$ (dashed curve); $\a = \a_0$, $\b= 2 \b_0$, $\eta=2 \eta_0$  (dotted  curve); and $\a = \a_0/2$, $\b=2 \b_0$, $\eta=2 \eta_0$ (solid curve, blue). Actions reducing $\ga$ by the same factor through action on the different parameters produce the same epidemic peak level, but with a substantially different dynamics.}\label{fig:asir2}
\end{figure}

In order to do this, we resort to numerical computations for a system with parameters strongly related to those of the SIR numerical computations considered in Sect. \ref{sec:episir} above, and along the same lines. The results of these are illustrated in Figs. \ref{fig:asir1} and \ref{fig:asir2}; these should be compared with Figs. \ref{fig:sir1} and \ref{fig:sir3}.

We also consider the quantities analogous to those considered in Table \ref{tab:sir}; these are given in Table \ref{tab:asir}.

\begin{table}
  \centering
  \bigskip
  \begin{tabular}{|c|c|c|c||l|l|r||r|r|r||}
  \hline
  $\a/\a_0$ & $\b/\b_0$ & $\eta/\eta_0$ & $\ga/\ga_0$ & $K_*/N$ & $I_*/N$ & $t_1$ & $t_2$ & $t_s$ & $\tau$  \\
  \hline
  1   & 1   & 1   & 1 & 0.730 & 0.058 & 29 & 28 &  94 &  82  \\
  1/2 & 1/2 & 1/2 & 1 & 0.730 & 0.058 & 58 & 55 & 188 & 164  \\
  2   & 2   & 2   & 1 & 0.730 & 0.058 & 14 & 14 &  47 &  41  \\
  \hline
  1/2 & 1   & 1   & 2 & 0.568 & 0.039 & 60 & 57 & 129 & 104  \\
  1   & 2   & 2   & 2 & 0.568 & 0.039 & 30 & 29 &  65 &  52  \\
  1/2 & 2   & 2   & 4 & 0.348 & 0.020 & 68 & 65 & 124 &  92  \\
  \hline
  \end{tabular}
  \caption{A-SIR model. Height and timing of the epidemic peak $K_* [t_1]$ and of the peak for symptomatic $I_*=I[t_2]$ (note $t_1 \not= t_2$) and time for reaching the ``safe'' level ($I(t_s) = 10^{-4} S_0$), together with duration of the interval $\tau = t_s -t_a$ defined in terms of symptomatic infectives, for the different combinations of parameters considered in the numerical runs of Fig.~\ref{fig:asir1} and \ref{fig:asir2}. See text. In all cases nearly all the population goes through infection, and a fraction $\approx \xi$ with symptoms.}\label{tab:asir}
\end{table}

It is rather clear that the same general phenomenon observed in the SIR framework is also displayed by the A-SIR equations. That is, acting only through social distancing leads to a lowering of the epidemic peak but also to a general slowing down of the dynamics, which means restrictive measures and lockdown should be kept for a very long time. On the other hand, less rough actions such as tracing contacts and the ensuing early isolation allow to reduce the peak without having to increase the time length of the critical phase and hence of restrictions.

\section{The ongoing COVID epidemic -- Italy}
\label{sec:ita}

Our considerations were so far quite general, and in the numerical computations considered so far we used constant parameter values which are realistic but do not refer to any concrete situation.

It is quite natural to wonder how these considerations would apply in a concrete case; we will thus consider the case we are more familiar with, i.e. the ongoing COVID epidemic in Italy.

The problem in analyzing a real situation is that the parameters are \emph{not}  constant: people get scared and adopt more conservative attitudes, and government impose restrictive measures. All these impact mainly on \emph{social distancing}, i.e. on the contact rate $\a$ \footnote{There are exceptions to this rule: e.g. in Veneto the regional strategy has been more focusing on early detection and prompt isolation, also thanks to contact tracing; with quite good results \cite{Crisanti}.}. Moreover, any measure shows its effect only after some delay, and of course not in a sharp way -- as incubation time is not a sharp constant but rather varies from individual to individual.

In related works \cite{Cadoni,Gasir} we have considered the SIR and the A-SIR models, and estimated parameters values -- that is, $\a$ and $\b$ for the SIR model, $\a, \b ,\eta$ and $\xi$ for the A-SIR one -- allowing them to satisfactorily describe the first phase of the COVID epidemic in Italy; we have then assumed that each set of restrictive measures has the effect of changing $\a$ to $r \a$ (with $0 < r < 1$ a reduction factor), and that this effect shows on after a time $\b^{-1}$ from the introduction of measures. This is a  very rough way to proceed, but it is in line with the simple approach of SIR-type modeling.

In Italy (total population $N \simeq 6*10^7$) a first set restrictive measures all over the national territory was  adopted on March 8, and another one on March 22. In the framework of the A-SIR model, our estimate of the parameters for the first phase of the epidemic (that is, before the first set of measures could have any effect, i.e. before March 15) was the following, with time measured in days:
$$ \a_0 \ \simeq \ 3.77 * 10^{-9} \ , \ \ \b^{-1} \ \simeq \ 7 \ , \ \ \eta^{-1} \ \simeq \ 21 \ ; \ \ \xi \ \simeq \ 1/10 \ . $$
See \cite{Gasir} for details on how these estimates are obtained.

When looking at data in the following time, we assumed that the contact rate $\a$ is reduced by the restrictive measures (and public awareness) and is changed from $\a_0$  into $\a_1 =r_1 \a_0$ between March 15 and March 29, and then into $\a_2 = r_2 \a_0$ from March 29 on (until measures will changed; at the moment it is expected that this should happen on May 4). Our best fit for the reduction factors $r_i$ is
$$ r_1 \ = \ 0.5 \ ; \ \ r_2 \ = \ 0.2 \ . $$

\begin{figure}
  \includegraphics[width=200pt]{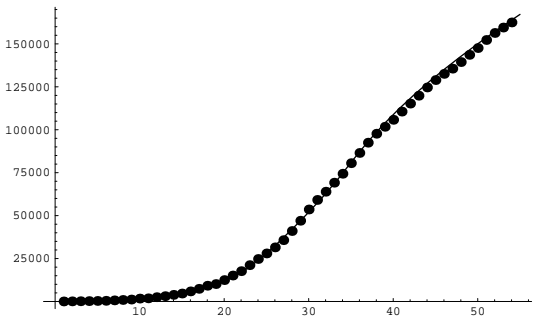}\\
  \caption{Fit of epidemiological data (points) by the function $R(t)$ arising from numerical solution of the A-SIR equations; see text for parameter values and other details. Here time is measured in days, with day 0 being February 20. Reproduced from \cite{Gasir}.}\label{fig:ITA}
\end{figure}

In Fig.~\ref{fig:ITA} we plot epidemiological data for the cumulative number of detected infections against the numerical solutions to the A-SIR equations for these values of the parameters, and for initial data obtained also from the analysis of real data. This shows a rather good agreement.

\subsection{Simple strategies}

We can now simulate further intervention, with four possible simple strategies \footnote{The general press does sometimes mention a different one, i.e. just relaxing restrictions with no other action; at the present stage of the epidemics, this would just ignite again the epidemic and is therefore non considered here.}:
\begin{enumerate}

\item Do nothing, i.e. keep the present social distancing measures;

\item Further reduction of $\a$  by a factor $\s$;

\item Increase of $\b$ and $\eta$ by a factor $\s$;

\item Simultaneous reduction of $\a$ and increase of $\b$ and $\eta$, all by a factor $\s$;

\end{enumerate}

We stress that these are not equally easy; we have considered the same reduction/increase factor in order to better compare the outcome of these three actions, but as so far all the intervention has been on $\a$, one should expect that further reducing it is very hard (even more so considering ``side effects'' of this on society, economics, and other sanitary issues  -- not to mention mental health issues \cite{Lancet}); on the other hand, no specific campaign designed to increase $\b$ has been conducted nationwide so far (with the exception of Veneto, as recalled in a previous footnote), so we expect action in this direction would be quite simpler an has more room for attaining a significant factor.

It should be mentioned, in this respect, that $\b^{-1}$ is already at its ``physiological'' level, just above the incubation time; so it can be further reduced only by contact tracing; the situation leaves more room for improving on $\eta$, as any campaign to identify asymptomatic infectives would reduce $\eta^{-1}$.

Despite these practical considerations, as already announced, we consider the same factor for the reduction of the contact rate and for the increase of the removal rates, and a homogeneous factor for the latter, in order to have a more direct comparison of the effects of these strategies.

\begin{figure}
  \begin{tabular}{|c|}
  \hline
  \includegraphics[width=200pt]{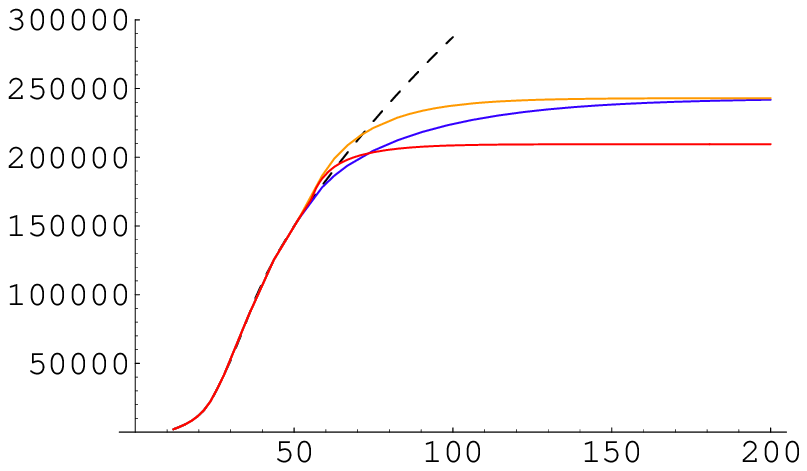}\\
  $R(t)$ \\
  \hline
  \includegraphics[width=200pt]{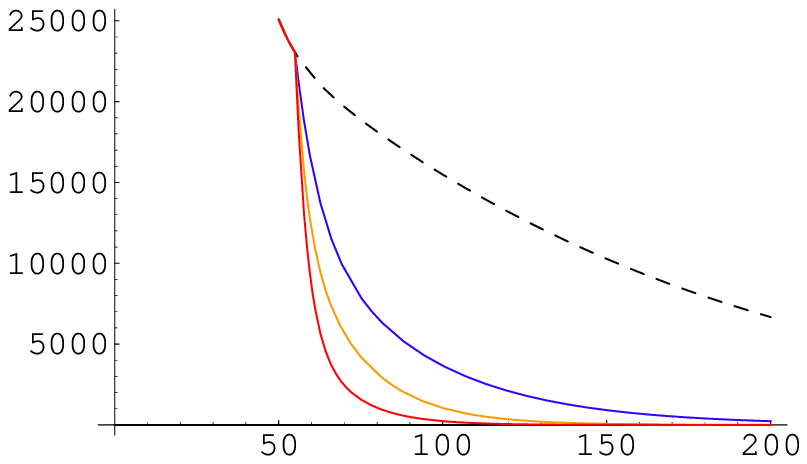}\\
  $I(t)$ \\
    \hline
  \includegraphics[width=200pt]{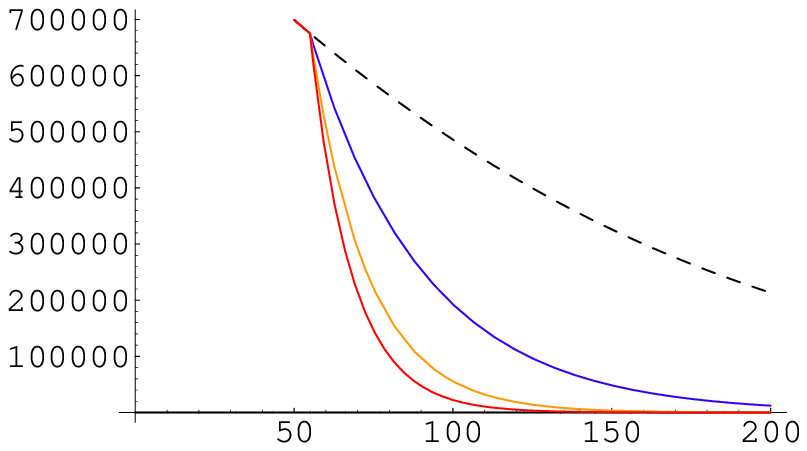}\\
  $K(t)$\\
    \hline
  \includegraphics[width=200pt]{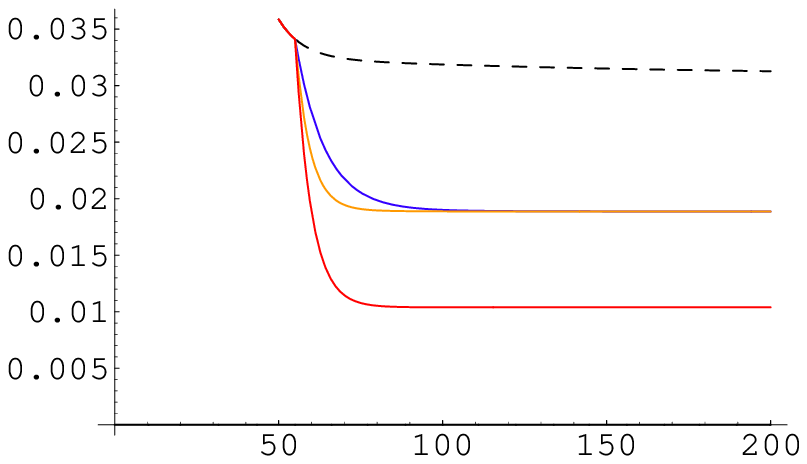}\\
  $x(t)$\\
    \hline \end{tabular}
  \caption{Simulation of different strategies after day 55 for Italy. Strategies correspond to those listed in the text, with factor $\s = 2$. The curves illustrate the predicted outcome under different strategies: no action (black, dashed); further reduction of $\a$ (blue); increase of $\b$ and $\eta$ (yellow); reduction of $\a$ and increase of $\b$ and $\eta$ (red). The plots represents, from the upper one to the lower one: cumulative number of symptomatic infected $R(t)$; number of symptomatic infectives $I(t)$; total number of infectives $K(t)= I(t)+J(t)$; ratio of symptomatic to total infectives, $x(t) = I(t)/K(t)$.}\label{fig:strat}
\end{figure}

We have then ran numerical simulations corresponding to the different strategies listed above; the outcome of these is given in Fig.~\ref{fig:strat}. Note that within each strategy the decay of $I(t)$ is faster than that of $K(t)$, and that the ratio between $I(t)$ ad $K(t)$, i.e. $x(t)=I(t)/K(t)$, is in all cases rather far from $\xi = 0.1$.

We also report in Table \ref{tab:meas} the relevant expected data for the time $t_s$ at which a safety level $I_s$ -- now assumed to be $I_s = 3,000$, i.e. half of the level reached when the first restrictive measures were taken -- is reached, the time interval $\tau = t_s -55$ from the day the new strategy is applied, and the total count $R_\infty$ of symptomatic infected. This table also reports the asymptotic value $x_\infty$ of the ratio $x(t)=I(t)/K(t)$ of symptomatic to total infectives; this is relevant in two ways: one the one hand one expects that only symptomatic infected may need hospital care, so a low level of $x$ means that albeit a number of infections will still be around only a small fraction of these will need medical attention, and COVID will not absorb a relevant part of the Health System resources; on the other hand, this also means that most of the infectives will be asymptomatic, i.e. attention to identifying and isolating them should be kept also when the number of symptomatic infections is very low. Note that for the strategy (1) what we report is not really an asymptotic value, but the value expected at December 31, as the decay is too slow.

In this regard, it may be interesting to note that according to our model at the end of April (day 70) we will have
\beq x \ \simeq \ 0.032 \ , \ \ \ \frac{R}{R+U} \ \simeq \ 0.124 \ . \eeq
Again according to our model and fit, at the same date the fraction of individuals having gone through the infection -- and thus hopefully having acquired long-time immunity -- would however be still below 3\% nationwide; this is obviously too little to build any group immunity. The situation could be different in the areas more heavily struck by the epidemic, such as Bergamo and Brescia.

\begin{table}
  \centering
  \begin{tabular}{|l||c|c|c||r|r|c|c||}
  \hline
   & $\a/\a_2$ & $\b/\b_0$ & $\eta/\eta_0$ & $t_s$ & $\tau$ & $R_\infty$ & $x_\infty$ \\
  \hline
  1. & 1   & 1 & 1 & 289 & 234 & $> 5*10^5$ & 0.031 \\
  2. & 1/2 & 1 & 1 & 107 &  52 & $2.4*10^5$ & 0.019 \\
  3. & 1   & 2 & 2 &  81 &  26 & $2.4*10^5$ & 0.019 \\
  4. & 1/2 & 2 & 2 &  68 &  13 & $2.1*10^5$ & 0.010 \\
  \hline
  \end{tabular}
  \caption{Time $t_s$ of reaching the safety level $I_s$, time from adoption of new strategy to $t_s$, and total final count $R_\infty$ of symptomatic infections, for the different strategies listed in the text, with a factor $\s = 2$.}\label{tab:meas}
\end{table}

\subsection{Discussion}

Figure \ref{fig:strat} and Table \ref{tab:meas} show, in our opinion quite clearly, that:
\begin{enumerate}

\item Continuing with present measures  with no accompanying action is simply untenable, as it would leave the country in this situation for  well over one further semester, and would also result in  a large number of new symptomatic infections -- hence also of casualties -- with the ensuing continuing stress on the Hospitals system.

\item Further restrictions in the direction of social distancing would have to be kept for nearly two further months; they would be effective in reducing the number of symptomatic infectives in the future. On the other hand, as individual mobility is already severely restricted, this would basically mean closing a number of economic activities which have been considered to be priorities so far (including in the most dramatic phase), which appears quite hard on social and economic grounds.

\item A campaign of early detection could be equally effective in reducing the number of infected and hence of casualties, but would require to maintain restrictions already in place for a much shorter time, less than one month. This should be implemented through contact tracing, which does not necessarily has to go through the use of technology endangering individual freedom, as shown by the strategy used in Veneto.

\item Combining further social restrictions and early detection would reduce the number of infections to a slightly smaller figure and would need to be implemented for  an even shorter time. This would however meet the same problems mentioned in item (2) above, albeit for a shorter time.

\item In all cases (except if no action is undertaken) the fraction of symptomatic infectives should soon fall to be between 1 and 2\%. This means we foresee a reduced stress on the Hospital system, but a continuing need to investigate, track and isolate asymptomatic infections, to avoid that at the end of restrictions they can spark a new epidemic wave.

\end{enumerate}

All in all, and repeating that we are not able to evaluate the social and economical cost of strategy 4, it appears that \emph{strategy 3},  i.e. raising the removal rates $\b$ and $\eta$, produces nearly optimal results with less strain on society and economy.

These considerations are of course not final,  also in view of the roughness built-in in the approach of SIR-type models; but we trust they offer a picture of what the consequences of different strategies are for what concerns strictly the epidemic dynamics. Decision makers will have of course to consider other aspects: social, economical, political, and also sanitary concerning other kind of pathologies.

Last but not least, tracing should be implemented in a way which is respectful of privacy and of individual freedom. We are of course not competent in discussing how this should be done in practice, hence will not discuss this point -- nor recall this matter later on in our discussion. We will just quote here Benjamin Franklin, who wrote that ``They who can give up essential Liberty to obtain a little temporary Safety, deserve neither Liberty nor Safety''.

\subsection{Modulating the best simple strategy}

It may be argued that reaching a factor 2 in the increase of $\b$ and $\eta$ may be too optimistic, and that imposing further social distancing measures may be non tenable socially. As for the second objection, we can only argue that -- in the case we best know, i.e. for Italy -- some of the imposed limitations may have been reasonable from a political point of view, but have very little epidemiological sense and could be safely removed \footnote{E.g. it is clear that forbidding single-personal physical exercise in isolated areas has no impact on fighting virus propagation. Similarly, people were led to amass in -- or in the waiting lines outside -- the open food shops by the prohibition to get out of home for other reasons that buy food, with sometimes grotesque exemplar cases. The general press is full of examples in this direction: old people being fined  for resting on a bench, or helicopters dispatched to fight a single man lying on the beach in complete isolation;  one may wonder how many lives could have been saved if all these energies would have instead been conveyed on checking what was happening in senior citizens residences, or more generally in surveillance of the medically effective distancing measures.}, while the contact rate could be more effectively lowered by strict regulations in workplaces, pushing the use of bike commuting instead of crowded public transport, or simply by making IPD widely available and their use enforced where it makes sense; this would reduce $\a$ without any social strain.

As for the first objection, it is difficult to say how far the increasing of the removal rates could be pushed, and in this sense the first anticipations in the general press of how the technological help for contact tracing would work (basically again awaiting that symptoms appears) are not very encouraging, as is the fact that only proximity of persons would be considered, and not the possibility of contagion through objects.

\begin{figure}
  \begin{tabular}{|c|}
  \hline
  \includegraphics[width=200pt]{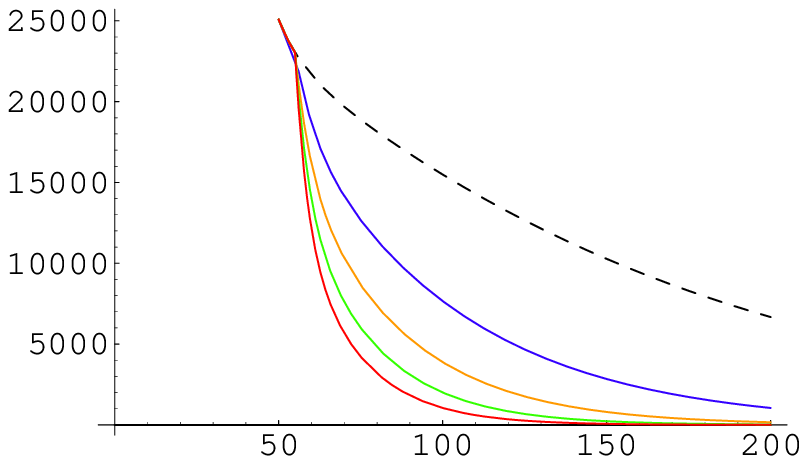}\\
  $I(t)$ \\
    \hline
  \includegraphics[width=200pt]{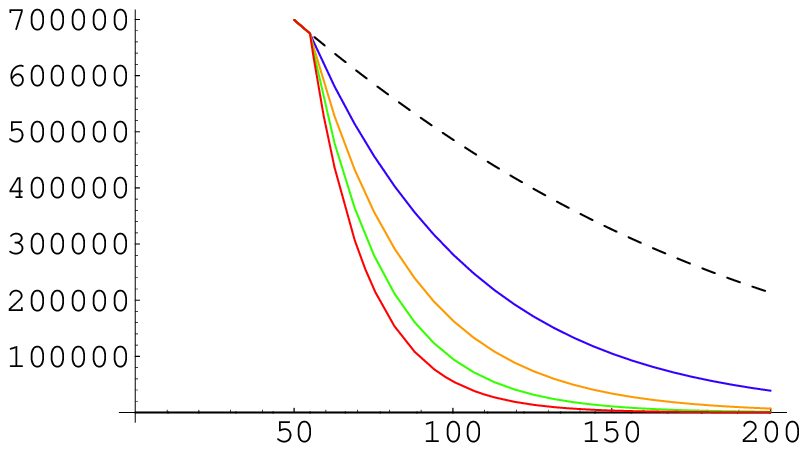}\\
  $K(t)$\\
    \hline \end{tabular}
  \caption{Simulation of strategy 3 with different factor $\s$ (see text) after day 55 for Italy. The curves illustrate the predicted outcome of strategy 3 with different factors: $\s=1$, i.e. no action (black, dashed); $\s=1.25$  (blue); $\s = 1.5$ (yellow); $\s = 1.75$ (green); $\s = 2.0$ (red). The plots represents the number of symptomatic infectives $I(t)$ and the total number of infectives $K(t)= I(t)+J(t)$.}\label{fig:modu}
\end{figure}

We have thus considered for the strategy 3 (increase of $\b$ and $\eta$ while maintaining $\a$ at its present level) different degrees of success, i.e. different values for the factor $\s$. The result of the simulations referring to this setting are summarized in Fig.~\ref{fig:modu} and in Table \ref{tab:modu}.

\begin{table}
  \centering
  \begin{tabular}{|l||r|r|r|r|r|}
  \hline
  $\s$ & 1.00 & 1.25 & 1.50 & 1.75& 2.00 \\
  \hline
  $t_s$  & 289 & 147 & 108 & 91 & 81 \\
  $\tau$ & 234 &  92 &  53 & 36 & 26 \\
  \hline
  \end{tabular}
  \caption{Time for reaching again a safe level of infected ($t_s$) and delay $\tau$ from the beginning of new strategy on day 55, for strategy 3 implemented with different factors $\s$. See text.}\label{tab:modu}
\end{table}

We see from Fig.~\ref{fig:modu} that albeit a moderate factor $\s$, e.g. $\s=1.5$ or even $\s=1.25$, produce a substantially faster decay of the number of infectives compared with the present situation, i.e. to the case where only social distancing is pursued. This is confirmed by the numerical values reported in Table \ref{tab:modu}.

Summarizing, the computations in this subsection confirm the outcomes of our discussion above, and actually strengthens them in that they show these are valid also if we do not succeed in greatly raising $\b$ and $\eta$, albeit of course the effect is stronger for higher raising factors $\s$.

\subsection{Relaxing the social distancing measures?}

One would hope that action on the removal time could allow to relax the social distancing measures. We have thus ran several numerical simulations as those shown in Fig.~\ref{fig:modu}, but with an $\a$ which is slightly higher than the present one; in particular, we have considered $\a = r * \a_0$ with $r = 0.35$, thus intermediate between that reached after the first restrictive measures and the present one (it is conceivable this could be reached going back to the first and milder restrictive measures but with a more efficient implementation of these).

Unfortunately, and as it had to be expected, this turns out to be possible without sparking a new epidemic growth -- nor slowing down the recovery from the present outburst -- only if the increase of $\b$ and $\eta$ outweigh the increase in $\a$, i.e. roughly speaking if $\ga = \b/\a$ is not lowered.

\begin{figure}
  \begin{tabular}{|c|}
  \hline
  \includegraphics[width=200pt]{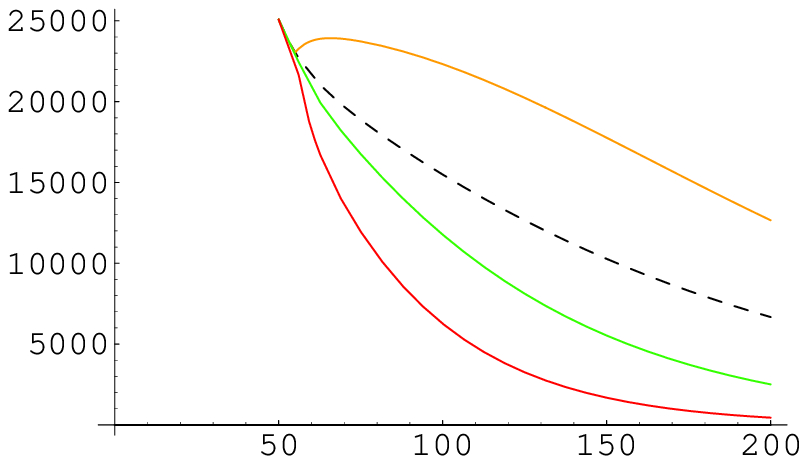} \\
  $I(t)/N$ \\
  \hline
  \includegraphics[width=200pt]{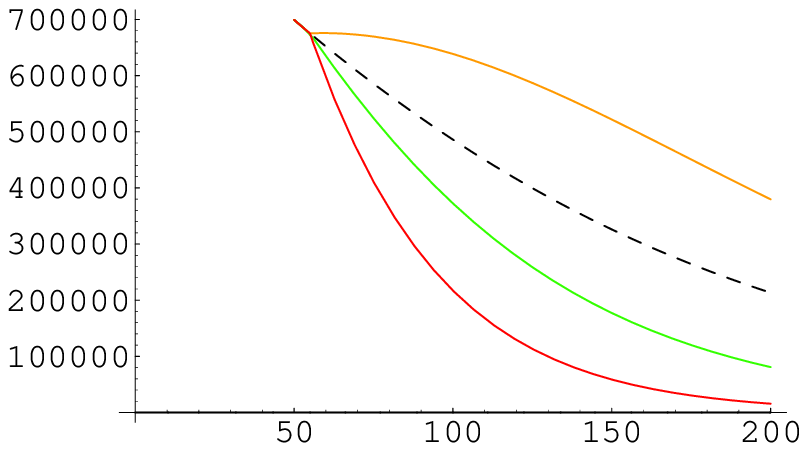} \\
  $K(t)/N$ \\
  \hline \end{tabular}
    \caption{Numerical simulation relaxing the contact rate reducing measures to $r=0.35$ after day 55 and simultaneously raising $b$ and $\eta$ by a factor $\s$ (see text). We plot the reference solution, i.e. no variation in $\a , \b , \eta$ (black, dashed) together with the solutions obtained for $\s= 1.5$ (yellow), $\s = 1.75$ (green) and $\s =2.0$ (red). The decay of infection is much slower than in Fig.~\ref{fig:modu}.}\label{fig:relax}
\end{figure}

We in particular have ran some numerical computations with $r=0.35$; these are depicted in Fig.~\ref{fig:relax}. Running the same simulations with $r=0.5$ -- that is, imagining one goes back to the contact rate reached after the first set of measures -- produces an increase of cases and a  large second maximum, with an epidemic peak about twice that of the first maximum.

\section{Conclusions}
\label{sec:conclu}

We have considered -- in the framework of ``mean field'' epidemiological models of the SIR type, hence disregarding any structure in the population -- how different strategies aiming at reducing the impact of the epidemic perform both in reducing the epidemic peak and the total number of people going through the infection state, and from the point of view of the time-span of the acute crisis state.

This has been discussed in general terms, both within the classical SIR model \cite{KMK,Murray,Heth,Edel,Britton} framework and with use of the recently formulated A-SIR model \cite{Gasir}, providing also some general results; in this setting, however, one deals with models with given parameters, constant in time.

On the other hand, in a real epidemic -- as the ongoing COVID one -- growing public awareness and governmental measures modify these parameters. We have considered a real case (Italy) from this point of view. After recalling that the models considered in this paper do quite well fit the epidemiological data so far, we have discussed what would be the impact of different strategies for the near future, showing that also in this case the model predicts a much shorter duration of the critical phase if further action concentrates on early detection of infectives rather than on social distancing, and this also if the shortening of the removal time is only quite small.

All in all, our model suggests something which experienced epidemiologists working in the field already know by direct experience \cite{Crisanti}. In the first wave, if the Health system is not ready to stand the epidemic wave or if the speed of the epidemics has been under-estimated \cite{GR0}, it is essential to slow down the contagion and social distancing, maybe in the form of a lockdown, is the simpler and faster way to achieve it.
After precious time has been obtained in this way, and when it comes to considering also the social and economic cost of a prolonged lockdown, the focus should shift to all possible means (respecting privacy and individual freedoms) of early detection and prompt isolation of infectives; so far the most effective means of achieving this is through tracing contacts, which may allow to isolate infectives before symptoms arise and thus to ``break the incubation time barrier''.




\end{document}